\documentclass[aps, prl, amsmath,amssymb]{revtex4-1}

\usepackage{amsmath,amssymb,graphicx}
\usepackage{epstopdf}

\begin{document}
\title{Micro-phase separation in two dimensional suspensions of self-propelled spheres and dumbbells}
\author{Clarion Tung$^1$, Joseph Harder$^1$, C. Valeriani$^2$ and A. Cacciuto$^1$}
\email{ac2822@columbia.edu}
\affiliation{$^{1}$Department of Chemistry, Columbia University\\ 3000 Broadway, New York, NY 10027\\ }
\affiliation{$^{2}$Departamento de Fisica Aplicada I, Facultad de Ciencias Fisicas, Universidad Complutense de Madrid, 28040 Madrid, Spain}
 
\begin{abstract}
\noindent  
We use numerical simulations to study the phase behavior of self-propelled spherical and dumbbellar particles
interacting via micro-phase separation inducing potentials. Our results indicate that 
under the appropriate conditions, it is possible to drive the formation of two new active states; a spinning cluster crystal, 
i.e. an ordered mesoscopic  phase having finite size spinning crystallites as  lattice sites, and a fluid of living clusters, i.e. a two dimensional fluid where each "particle" is a finite size living cluster. We discuss the dynamics of these phases and suggest ways of
extending their stability under a wide range of active forces.
\end{abstract}

\maketitle

\section*{Introduction}

Spontaneous pattern formation is a ubiquitous phenomenon in nature and arises in both equilibrium and out-of-equilibrium systems. Apart from the many biological examples~\cite{Alberts,Meinhardt}, important  synthetic systems such as colloids and block copolymers have been shown to exhibit complex spatial equilibrium patterns upon self-assembly. Control of patterns at the micro and  nanoscale is integral to the development of materials with novel optical, electrical, and rheological properties~\cite{Vedmedenko}. 

One route to spontaneous pattern formation in equilibrium systems is achieved via micro-phase separation~\cite{Leibler}, a phenomenon that typically occurs when geometrical or chemical constraints  prevent a system from fully phase separating.  Block copolymers, for instance, exhibit a wide variety of patterns upon micro-phase separation that can be controlled by tuning the relative length of the two blocks, from lamellae to cylinders to networks \cite{Blanazs, battaglia, Hinsberg}.  An alternative route is achieved with competing interactions. In colloidal supensions, for instance, the interplay between a short range attraction and a long range repulsion  has been shown to lead to micro-phase separation into a variety of patterns~\cite{poon2} with 
symmetry dependent on the relative weight of the two interactions \cite{Sciortino,imperio2004bidimensional, imperio2006microphase, imperio2007microphase, imperio2008rheology}. In these cases, 
the short range attraction is usually induced by depletion, hydrophobic or van der Waals forces, while the long range repulsion may come from dipolar forces or screened electrostatics \cite{de2000dipolar, muratov2002theory, ghezzi1997formation, ghezzi2001pattern}.  
%Understanding the mechanisms underlying the self-organization of nanocomponents over a wide range of times and length scales is a critical step towards a bottom-up  approach to materials design via the process of self-assembly
%Along with equilibrium systems, active systems are also capable of exhibiting dynamic, steady-state patterns.  
%Multiple studies have observed phase separation in active matter systems, in which active colloids form large clusters that span the system \cite{buttinoni2013dynamical, fily2012athermal, fily2014freezing, lobaskin2013collective}.  

Recent experiments in two dimensions have shown that dilute suspensions of self-propelled colloidal particles can self-assemble into ``dynamic'', ``living'' crystals \cite{palacci2013living,theurkauff, buttinoni2013dynamical}, where finite-sized aggregates continually join, break apart, dissolve, and reform.  
In this case, what limits the growth of the crystal to a macroscopic size $-$ the scenario that would be favored due to inter-particle attractive interactions $-$ is the self-propulsion of the active particles. This behavior seems to be specific to spherical particles, as these are able to freely re-orient within the developing 
crystallites under the influence of thermal forces, thus creating
large stresses within the crystal. 
The formation of living clusters has also  been recently observed in computer simulations of a three dimensional diluted suspension of self-propelled attractive spheres~\cite{mognetti2013living,filion}; however, in three dimensions clusters did not show any crystalline order. 
Experiments with swimming bacteria in a three dimensional container and in the presence of small polymer depletants~\cite{schwarz2012phase} reported the formation of 
micro-clusters with net rotational velocity dependent on the size and shape of the clusters (micro-rotors). This 
was also observed in simulations of two dimensional dumbbells with different motilities~\cite{maren}.
Both living crystals and micro-rotors are beautiful examples of active finite-size structures. 
However, it is not clear whether a fluid with living crystals represents a truly stable phase~\cite{PRXX}; 
furthermore, the  micro-rotors observed experimentally are short-lived, and  over time they merge to phase separate into macroscopic structures.

The aim of this paper is two-fold. On one hand, we want to understand whether 
explicitly introducing a micro-phase separation-inducing long range potential between self-propelled attractive particles 
 can be exploited to control size and activity of living crystals and micro-rotors.
On the other, we seek to understand how self-propulsion of  colloidal particles can be used as a means to alter the morphological features of the micro-patterns formed by the passive  particles.
With this in mind, we explore the behavior of two dimensional dilute suspensions of active Brownian spheres and dumbbells interacting with competing short and long range potentials for different degrees of particle activity. We report a rich and complex range of physical behavior
that includes the formation of mesophases of spinning and living crystals.

 \section*{Methods}
We study systems of \(N\) active colloidal spheres and dumbbells in a  two dimensional box of size $L$ with periodic boundary conditions.  Spherical particles are characterized by a diameter \(\sigma\), and dumbbells are modeled as two spheres, each of diameter $\sigma$, rigidly connected to each other at contact distance, with length $\l=2\sigma$.  Both spheres and dumbbells   undergo Brownian motion, with an additional constant propelling force \(F_a\) acting along an orientation vector 
\(\boldsymbol{n} = (\cos \theta, \sin \theta )\); for dumbbells this axis coincides with the particle long axis.  The equations of motion of these systems follow Brownian dynamics:

\begin{eqnarray}
\boldsymbol{\dot{r}_i} =\frac{D}{k_{\rm B}T}\left(-\nabla_i V(r) + |F_a|\,\boldsymbol{n_i}\right) + \sqrt{2D}\,\boldsymbol{\xi_i}  \\  
\boldsymbol{\dot{n}_i} = -\frac{D_r}{k_{\rm{B}}T} \nabla_{\boldsymbol{n_i}} V(r)\, + \sqrt{2D_r}\
\,\boldsymbol{\xi_{r_i} } \times \boldsymbol{n_i},
\end{eqnarray}
where the first equation describes  the translational  and  second  the rotational motion of the particles. $V(r)$ is the interparticle potential.
The first term on the right side of Eq.~2 accounts for the external torques that develop exclusively for dumbbellar interactions. The relation between the translational, $D=k_{\rm B}T/\gamma$, and the rotational diffusion coefficient, $D_r$, for spherical particles is set to  \(D_r = (3D)/\sigma^2 \).
Here $\gamma$ is the friction coefficient and is set to 1.
Following~\cite{heptner2015equilibrium}, the translational dynamics of dumbbells is solved independently for the 
parallel ($\bf{r_{i\,/\!/}}$) and the perpendicular  ($\bf{r_{i\perp}}$) components of the particles' coordinates 
with respect to the active axis $\boldsymbol{n_i}$.  Correspondingly, there are
 two different diffusion coefficients $D\rightarrow (D_{\!/\!/},D_{\perp})$ 
with $D_{\perp}=D_{\!/\!/}/2$ and \(D_r = (6D_{\!/\!/})/l^2\)~\cite{kirkwood}.  
The solvent induced random fluctuations for the translational $\xi_i(t)$ and rotational $\xi_{r_i}(t)$ degrees of freedom obey the relations \(\langle\xi_i(t)\rangle = 0\) and \(\langle\xi_i(t)\cdot \xi_j(t)\rangle = \delta_{ij}\delta(t-t')\) and \(\langle\xi_{r_i}(t)\rangle = 0\) and \(\langle\xi_{r_i}(t)\cdot \xi_{r_j}(t)\rangle = \delta_{ij}\delta(t-t')\), respectively.

Following Imperio et al.~\cite{imperio2007microphase}, we set the interaction between any two non-connected spheres in the system to be
\begin{eqnarray}
\label{IRpot} 
V(r) = \varepsilon \left( A\left(\frac{\sigma}{r}\right)^n - \frac{\epsilon_a \sigma^2}{R_a^2}\exp\left(-\frac{r}{R_a}\right) + \frac{\epsilon_r \sigma^2}{R_r^2}\exp\left(-\frac{r}{R_r}\right) \right) 
\end{eqnarray}
The interaction parameters are \(n=12\), \(A=0.018\), \(R_a=\sigma\), \(\epsilon_a=1\), \(R_r=2\sigma\), and \(\epsilon_r=1\), and 
$\varepsilon$ ranging from 25 to 50 $k_{\rm B}T$.  This set of parameters has been chosen 
to guarantee micro-phase separation of the passive system as  reported in~\cite{imperio2007microphase}.
This potential consists of a Lennard-Jones  repulsion at short distances  to enforce excluded area (the first term); a short range attractive well to induce clustering of the particles (the second term); a long range soft repulsion to guarantee micro-phase separation(the last term). The potential is cut off and truncated at \(r=10\sigma\).

Throughout this study, we used dimensionless units, where lengths, energies and times are expressed in terms of $\sigma$, $k_{\rm B}T$ and   $\tau=\sigma^2/D$ (for spheres),  $\tau=\sigma^2/D_{\!/\!/}$ (for dumbbells), respectively.
We use the dimensionless P\'eclet number, defined as $\mathrm{Pe}=\frac{|F_a|D}{k_{\rm B}T}\frac{\tau}{\sigma}$ for spheres 
and $\mathrm{Pe} = \frac{|F_a|D_{/\!/}}{k_{\rm B}T}\frac{\tau}{\sigma}$ for dumbbells, as 
a measure of the degree of self-propulsion of the particles.
The time step was set to \(\Delta t=5\times10^{-5}\tau\) and all simulations were run for a minimum of $10^7$ time steps, with  $N=1000-3000$ spheres, and $N=500-2000$  dumbbells, at area fractions  \(\phi=\frac{\pi}{4}N\sigma^2/L^2\) ranging from 0.05 to 0.5, and P\'eclet numbers from \({\rm Pe}=0\) to \({\rm Pe}=40\). 
 
\section*{Results}
 
For the chosen parameters of the interaction potential, in the absence of active forces (Pe=0) and at small area fractions (typically smaller than $\phi\simeq 0.5$), both spheres and dumbbells   micro-phase separate to form cluster crystals (see Fig. 2 and 6), i.e. isotropic finite-sized crystallites of particles acting as lattice sites of a larger mesoscopic phase with overall hexagonal packing $-$ apart from the dislocations that develop as a result of the size polydispersity of the small crystalline clusters.  
As the area fraction increases, so does the average size of the  crystallites, until they eventually merge to form linear aggregates of finite width. This phase behavior is consistent with what reported in previous studies~\cite{imperio2007microphase}.
Because spheres can freely (at no energy cost) rotate within a crystallite without affecting its underlying structure, but dumbbells cannot, aggregates made of spheres and dumbbells behave quite differently when particle self-propulsion is switched on. We shall therefore discuss the two cases separately. 

\subsection*{Spheres}
\begin{figure} [h!]
\centering
%\begin{minipage}{.45\textwidth}
  \centering
  \includegraphics[width=0.5\linewidth]{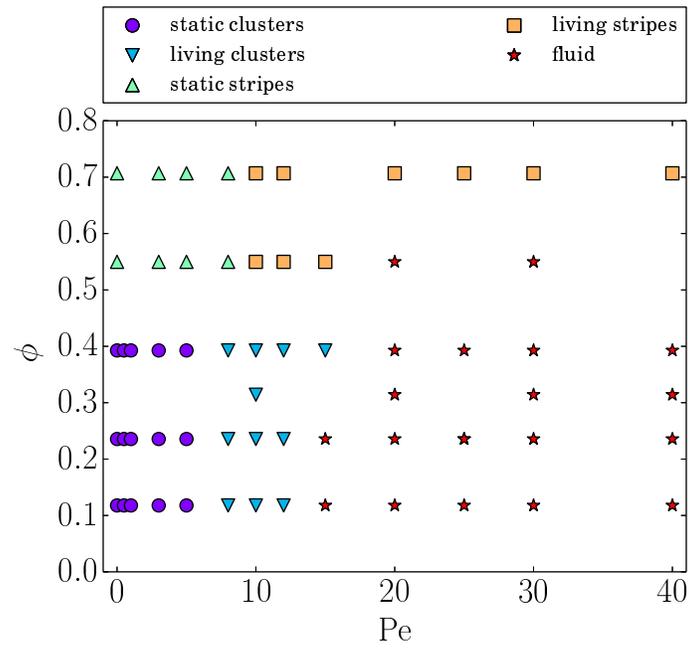}
  \caption{A summary of all the structures observed in our simulations of active spheres as a function of P\'eclet number \(\mathrm{Pe}\) and area fraction \(\phi\), at \(\varepsilon=25k_{\mathrm{B}}T\). }
  \label{pd} 
%\end{minipage}%
\end{figure}

\begin{figure} [h!]
%\begin{minipage}{.45\textwidth}
  \centering
  \includegraphics[width=0.5\linewidth]{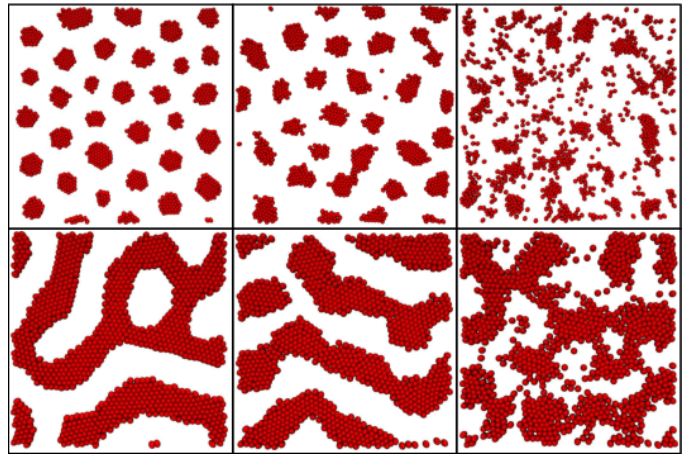}
  \caption{Self-propelled spheres. Top: snapshots at \(\phi = 0.24\), bottom: \(\phi = 0.55\) (See Fig. 1 as a reference).  
  From left to right: \(\mathrm{Pe}=0, 10, 20\).   At \(\mathrm{Pe}=10\), meso-phases of living clusters are observed at lower densities, and living stripes are seen at higher densities.}
  \label{diskspic}
%\end{minipage}
\end{figure}

All simulations with the spheres have been performed at
$\varepsilon=25 k_{\rm B}T$. This corresponds to a pairwise binding energy of $\varepsilon_0\simeq -5 k_{\rm B}T$; this particular  value was selected to coincide with previous numerical studies with this potential~\cite{imperio2007microphase}.
A structural diagram in terms of the area fraction and P\'eclet number is shown in Fig.~\ref{pd}, while  Fig.~\ref{diskspic}  shows snapshots of the corresponding morphologies taken from our simulations.

% At low volume fractions, the phase formed by the passive particles remains largely unaffected for small P\'eclet numbers, but as soon as Pe becomes larger than $\varepsilon_0/k_{\rm B}T$, at first single particle exchange  between  previously isolated (static) clusters begins to occur, this behavior resembles that 
% observed in cluster crystals formed by soft particles compressed at high densities~\ref{QQQQQQ}, but upon further increase of Pe, ever larger "chunks" of material are exchanged between the clusters. This occurs because individual crystallites begin to locally split and reassemble under the stresses imposed by the active forces (living crystals~\ref{Chainkinnnnnnnnn}).

At low area fractions and small P\'eclet numbers, we observe the expected mesoscopic crystal of static clusters.  As soon as Pe becomes larger than \(|\varepsilon_0|/k_{\mathrm{B}}T\), not only does particle exchange between clusters begin to occur, resembling the behavior observed in cluster crystals of soft particles at high densities \cite{moreno2007diffusion,mladekk}, 
but clusters also begin to deform, split, and merge with neighboring clusters under the stress imposed by the active forces of the constituent particles.   The net result is the formation of a new phase consisting of a fluid of living clusters. In this phase, the motion of the clusters away from the lattice sites of the mesophase is driven by the "living" character of the clusters themselves. 
In fact, as soon as a cluster splits and one or both of its parts are incorporated into 
the neighboring clusters, an imbalance in the long range forces holding together the mesophase takes place, resulting in
a global rearrangement of the clusters or in their further repartitioning to re-establish the overall balance of the forces. 
Movies of this behavior are shown in the SI.

\begin{figure}[h]
\centering
%\begin{minipage}{.45\textwidth}
  \centering
  \includegraphics[width=0.5\linewidth]{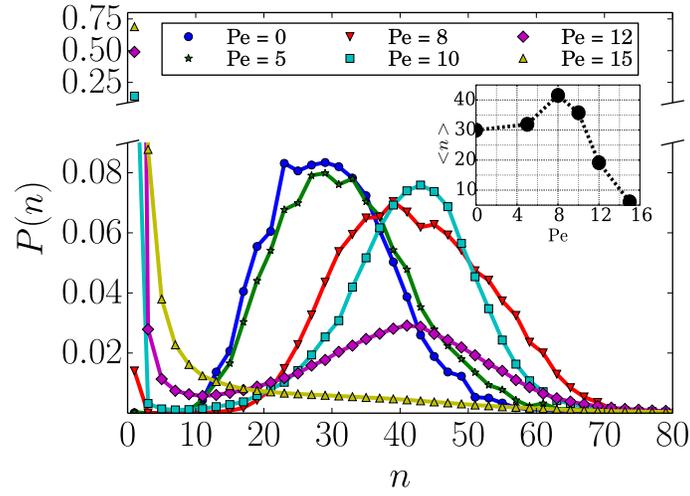}
  \caption{Cluster size distributions as a function of \(\mathrm{Pe}\) are shown for spheres at \(\phi=0.24\), with \(\varepsilon=25k_{\mathrm{B}}T\).  At \(\mathrm{Pe} \simeq 5\), fluids of living clusters are seen, for which particle exchange occurs.  At \(\mathrm{Pe}\simeq 15\), there is no longer a stable cluster size and the system is a fluid.  The inset shows the average cluster size for increasing \(\mathrm{Pe}\).   }
  \label{clustersize}
%\end{minipage}%
\end{figure}

For small values of  \(\mathrm{Pe}\), the mesophase is able to relax to accommodate the vacancies or density inhomogeneities
created by the living clusters.  But at large values of  \(\mathrm{Pe}\), the rate of cluster subdivision becomes too fast, the meso-phase relaxation cannot keep up and the mesophase finally melts,  resulting in a  system of smaller clusters in a gas of dispersed active particles (fluid).    
This behavior can  be easily tracked  in the steady state cluster size distributions $P(n)$ shown in Fig.~\ref{clustersize}, where
we study how $P(n)$ changes when the system moves from the static cluster, to the living cluster, and all the way to the fluid phase, by changing the value of Pe. The well defined distributions centered around large values of $n$ for small P\'eclet numbers give rise to a large peak at $n=1$ (isolated particles) and a very wide distribution of clusters sizes as the meso-phase melts.
Also notice that the average size of the clusters $\langle n\rangle$  (see inset of Fig.~\ref{clustersize}) tends to increase  upon activating the self-propulsion of the particles, but $\langle n\rangle$ shows an overall non-monotonic behavior with the strength of the active forces. 
Our data indicate that the decrease of the average size for large activities is followed by a significant  broadening of the cluster's polydispersity, and melting of the mesophase.  

\begin{figure}[h]
%\begin{minipage}{.45\textwidth}
  \centering

  \includegraphics[width=0.5\linewidth]{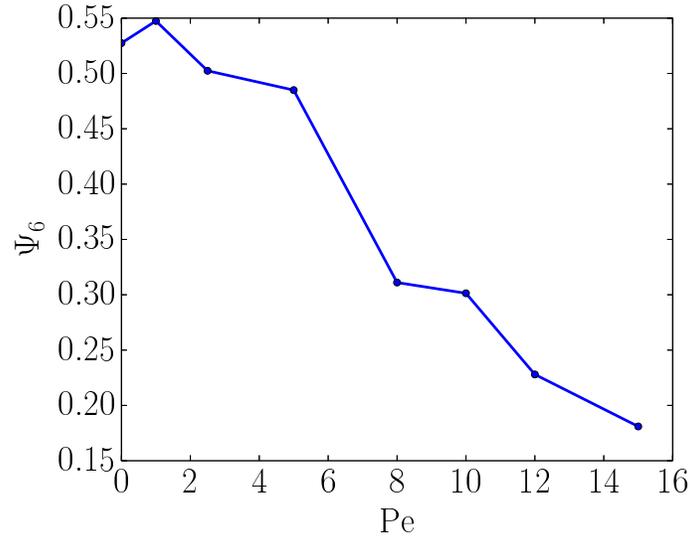}
  \caption{Cluster-to-cluster bond order parameter \(\Psi_6\) as  a function of \(\mathrm{Pe}\) for \(\phi=0.24\) of the mesophase formed 
  by active spherical particles. }
  \label{q6plot}
%  \end{minipage}

\end{figure}

To quantify the order of the mesophase as a function of Pe, we also tracked the cluster-to-cluster  bond order parameter  \(\Psi_6\) \cite{nelson1979dislocation}.  
In our simulations, cluster connectivity is obtained using a simple Delaunay triangulation of the clusters' centers of mass, and \(\Psi_6\) is defined as:
\[ \Psi_6 = \frac{1}{N_c}\sum_{i=1}^{N_c}\,\frac{1}{N_b(i)} \sum_{j=1}^{N_b(i)} e^{\mathbf{i} 6\theta_j}\,, \]
where \(N_c\) is the total number of clusters in a configuration, \(N_b(i)\) is the number of  clusters connected to cluster $i$, and \(\theta_j\) is the angle between a neighboring cluster and a reference axis.
As expected (see Fig.~\ref{q6plot}),   \(\Psi_6\) shows the deterioration of order of the meso-phase with increasing Pe.

An analogous behavior occurs at higher densities, where the passive phase is characterized by linear aggregates, or stripes, spanning the system size. In this case, intermediate propelling forces lead to the
continuous recombination of the maze formed by the  particles via formation and breakage of junctions between the stripes. At even larger \(\mathrm{Pe}\) the stripes begin to lose connectivity and break up, giving
rise to large-scale density fluctuations.

 \subsection{Dumbbells}
\begin{figure}
\centering
%\begin{minipage}{.45\textwidth}
  \centering
  \includegraphics[width=0.5\linewidth]{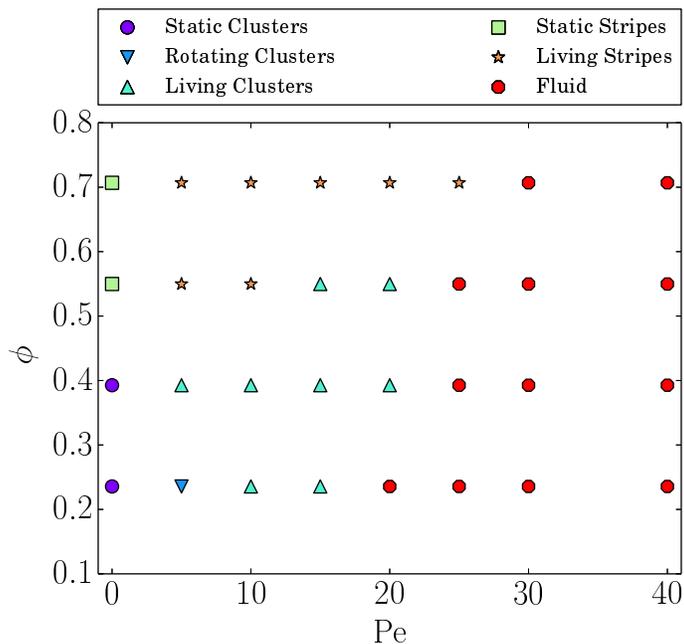}
  \caption{A summary of all the structures observed in our simulations of active dumbbells as a function of P\'eclet number \(\mathrm{Pe}\) and area fraction \(\phi\), at \(\varepsilon=25k_{\mathrm{B}}T\).  In contrast to the spheres, at low values of Pe, clusters rotate, each with a specific handedness, and translate.}
  \label{pd_rods}
%\end{minipage}%
\end{figure}

\begin{figure}
%\begin{minipage}{.45\textwidth}
  \centering
  \includegraphics[width=\linewidth]{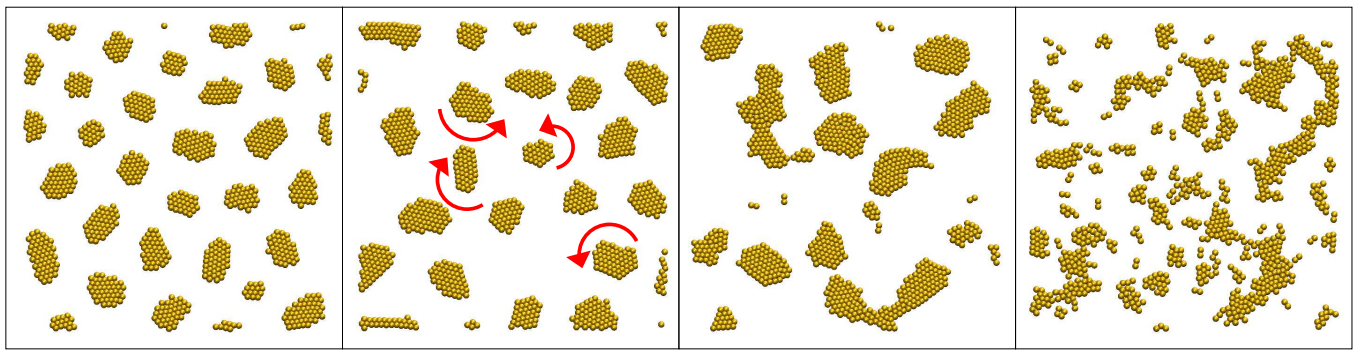}
  \caption{Dumbbells at \(\phi=0.24\) showing the variety of phases formed.  From left to right: Pe = 0 (static clusters), Pe = 5 (rotating clusters), Pe = 15 (living clusters), Pe = 40 (fluid phase).}
  \label{rodspic}
%\end{minipage}
\end{figure}

We also study the behavior of dumbbells for \(\varepsilon=25k_{\mathrm{B}}T\) and over the same range of activities and area fractions as the spheres.
Fig.~\ref{pd_rods} shows the structural diagram of the different regions observed 
in our simulations in terms of the area fraction and the P\'eclet number, and Fig.~\ref{rodspic} shows corresponding snapshots from simulations.
The phase behavior of the passive system is analogous to that observed for spherical particles, 
where mesocrystals of hexagonally packed finite size clusters are observed at low area fractions,  and stripes develop for  sufficiently large densities. 
In contrast to the spheres, however, when self-propulsion is switched on at low values of Pe and at low densities,  clusters acquire a well defined rotational motion and translational motion that develops as a result of the nonzero internal net force acting on the clusters of dumbbells, as discussed in~\cite{schwarz2012phase,maren}. Thus, the first phase that one encounters at low area fractions and at small activities is a fluid of rotating and translating compact clusters.
As observed in our simulations with  spherical particles, in this regime, an increase of the particle activity leads to an overall increase of the average size of the clusters. 
Further increase of Pe takes the system into a different state, and the rotational and translational velocity of the clusters increases. 
This state is characterized by a fluid of drifting, rotating, and living clusters. Shear forces within the clusters are now 
able to break them apart, and their net translational velocities are strong enough to make whole clusters collide and merge. 
For this range of values of Pe, single dumbbells rarely leave the clusters individually and mostly rearrange around the surface of
the individual clusters, leading to a less well defined rotational velocity.
Finally, for even larger forces, single dumbbells are able to fly off the clusters, which melt and 
reduce in size, as observed in the case of the spheres.

\begin{figure}[h]
\centering
%\begin{minipage}{.45\textwidth}
  \centering
  \includegraphics[width=0.5\linewidth]{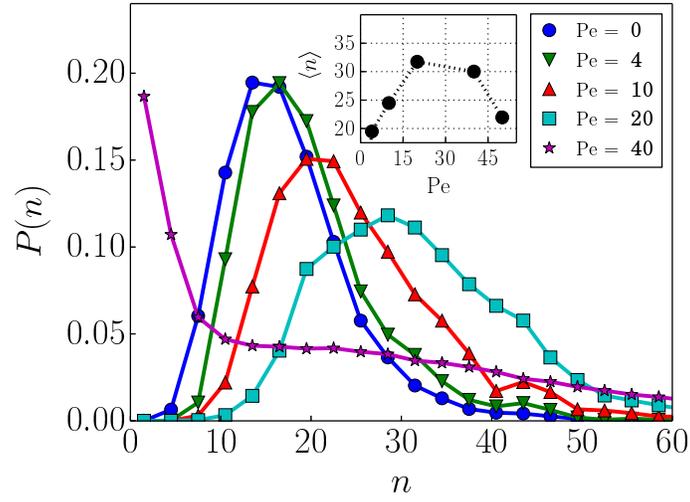}
  \caption{Cluster size distributions for dumbbells at \(\phi=0.24\) and \(\varepsilon = 50 k_{\rm{B}}T\) with increasing Pe.  The cluster distribution shifts towards larger clusters and broadens with increasing activity until the system becomes a fluid. The inset shows the average cluster size for increasing values of  Pe.  }
  \label{rod_cluster}
%\end{minipage}%
\end{figure}

The linear, stripe-like aggregates of dumbbells found at high area fractions are similar to those observed for spheres, and living stripes are also seen 
at high activity.  In contrast to the dynamic maze-like patterns observed for spheres, the morphology of the living stripes of dumbbells 
changes through large portions of stripes shearing off and fusing with neighboring stripes.  This is due to the large net internal forces 
that continually shear and break the stripes.
At even larger activities (not shown), clusters  can easily completely melt and reform
resulting in a phase characterized by large-scale density fluctuations as 
observed in active spheres at large densities and activities.
Fig.~\ref{rod_cluster} and its inset show the cluster size distributions at low densities for different values
of Pe, and the non monotonic behavior of the average cluster size with degree of activity. 
 
\begin{figure}[h] 
%\begin{minipage}{.45\textwidth} 
  \centering
  \includegraphics[width=0.5\linewidth]{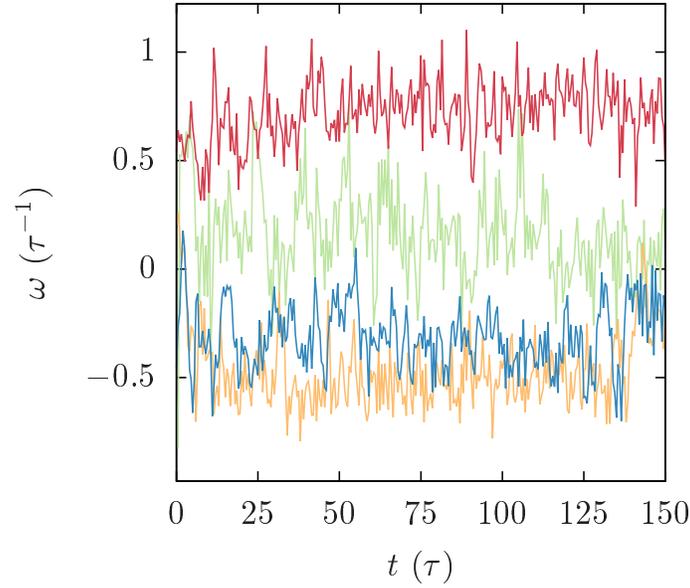}
%   \caption{ $\varepsilon=0.1$ for these guys, $\rho=0.3$, $t=5*10^{-5}$ .  Inset shows the rotational autocorrelation function for 20 trajectories at \(F_a=10\).}
  \caption{ The angular velocities for several rotating dumbbellar clusters containing roughly 20-30 dumbbells from a system at \(\phi = 0.24\), \(\varepsilon = 50 k_{\rm{B}}T\), and Pe = 20 are shown.  The clusters show unidirectional rotation over the duration of the simulation.}
  \label{rotvel}
%\end{minipage}
\end{figure}

\begin{figure} [h!]
  \centering
  \includegraphics[width=0.5\linewidth]{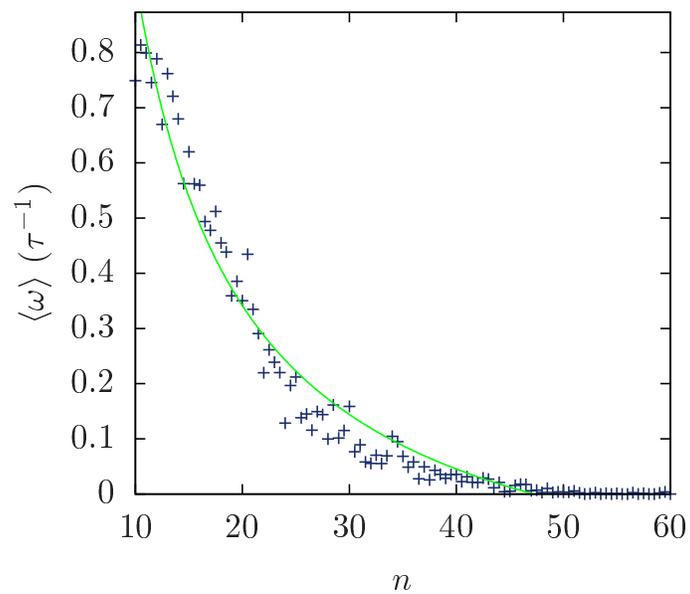}
  \caption{Average rotational speed $\langle \omega \rangle$ of a cluster of dumbells of size $n$, as measured in our simulations, for \(\phi = 0.24\), \(\varepsilon = 50 k_{\rm{B}}T\), and Pe=20.  The solid line is the expected theoretical dependence $\langle \omega \rangle\sim 1/n$.}
  \label{rotdecay} 
\end{figure}

Fig.~\ref{rotvel} shows the angular velocity of a few selected clusters containing roughly thirty dumbbells in the phase of rotating clusters at
low density/low Pe over time. This indicates that the rotational motion in this regime persists over time, but the specific direction and velocity 
of this rotational motion varies from cluster to cluster and depends on the specific arrangement of the dumbbells 
constituting each crystallite.  The simulations to measure the angular velocity of the rotors are performed 
with \(\varepsilon = 50k_{\mathrm{B}}T\) to extend the stability of this phase against the active forces and better observe its phenomenology.     
Overall, the angular velocity is expected to decrease with the size of the clusters  as previously reported in 
~\cite{schwarz2012phase,maren}. A quick estimate of the angular velocity dependence on
the cluster sizes can be obtained by assuming that the axes of $n$ dumbbells forming a cluster of radius $R$, are randomly oriented.
The sum of these random forces  will add up to a net force imbalance  $f\simeq F_a\sqrt{n}\simeq F_aR$. Thus, one can expect 
an average torque acting on the cluster for any given random configuration of dumbbells that scales as $\tau=fR\simeq F_a R^{2}$. Balancing this torque with the rotational friction that we take 
to be proportional to  $R^4\omega$, where $\omega$ is the angular velocity (this is the result for typical simulations without hydrodynamics~\cite{maren}), one expects $\omega\sim F_a/R^2\simeq F_a/n$. Fig.~\ref{rotdecay} shows the dependence of the average angular velocity $\omega$  on cluster size for our dumbbell rotors. Our simulation data agrees with our analysis and shows that the average angular velocity decreases with cluster size. 
 
One of the most exciting new phases for dumbbells is the fluid of rotating clusters observed
at low densities and small P\'eclet numbers. These cannot form with spherical particles as spheres can freely rotate
within the finite size clusters, but dumbbells become sterically locked in place by the attractive forces. As a result, 
once a cluster  of dumbbells is formed, that cluster will acquire a constant angular velocity over time.
As mentioned above, the stability of the clusters 
can be enhanced by increasing the strength of the potential \(\varepsilon\), however,
the overall order of the meso-phase is disrupted because of their active translational motion.
This occurs because, as explained above, for any configuration of $n$ randomly assembled dumbbellar clusters, a net imbalance of forces proportional to $F_a \sqrt{n}$  will on average develop in a random direction.
However, one could envision that increasing the repulsive part of the 
interaction between the particles could create large energy barriers between the clusters, and in this way, limit the
active translation of the clusters (at low Pe), but still keep their rotational motion. This would lead to the formation
of yet another phase characterized by a cluster crystal of with spinning lattice sites. Our simulations 
show that this is indeed the case, and can be achieved by setting
$\varepsilon_r=2.5\varepsilon_a$ in the potential $V(r)$.
Simulations for spheres with these parameters show better defined clusters, but the splitting and recombination of the clusters $-$ which now occurs at larger P\'eclet numbers $-$ leads again to mass transfer to the neighboring sites and subsequent relaxation
of the whole mesophase. A movie of the dumbbellar spinning cluster crystal can be found in the SI.

\section*{Conclusions}

In this paper we studied the phase behavior of diluted suspensions of self-propelled 
dumbbells and spheres interacting with a micro-phase separation inducing potential.
Specifically, we considered a pair potential composed of a short-range attraction and a long-range repulsion.
This is a standard potential that has been used  extensively in the literature of micro-phase separation
of colloidal particles and is one that mimics the interaction of weakly charged particles in the presence of depletants. Both forces are easily tuned by either changing the salt concentration in solution or the density of the depletants. 

Our results indicate that for a range of parameters, it is possible to induce
the formation of two previously unobserved states; a spinning cluster crystal, i.e. an ordered mesoscopic  phase having finite size spinning crystallites as  lattice sites, and a fluid of living clusters, i.e. a two dimensional fluid where each "particle" is a finite size living crystallite.
The first state develops from the self-assembly of dumbbells, whereas the second state occurs for spherical particles. We suggest ways to increase the stability of these states by appropriately selecting the relative
weight of the two competing interactions with respect to the self-propulsion.

Several groups have observed  phase separation of self-propelled hard particles at high activity and area fraction
(see for instance~\cite{sp1,sp2,sp3}).   To understand whether a sufficiently large self-propulsion would
overwhelm the role of the repulsive interactions and lead to a re-entrant behavior as observed in colloidal particles 
with attractive interactions~\cite{redner}, we performed simulations for Pe up to 150 and area fractions up to 0.5.  We did not observe
a complete phase separation of the system, suggesting that at least within this range of parameters, the long range repulsive tail of the potential 
prevents this transition.

Another interesting result that emerges from our simulations concerns the size of the clusters that
form at low densities. For both spherical and dumbbellar particles, we find that the average size of the crystallites varies non-monotonically with the strength of the propelling forces. While preparing this manuscript, this result was also reported in~\cite{mani2015effect} for the case of spherical particles.

Although our system is two dimensional, the combination of short-ranged attractive and long-ranged repulsive interactions also leads to microphase separation in three dimensions~\cite{Sciortino}.  We speculate that both the fluid of living clusters and crystal of micro-rotors will survive, but the morphological properties of the clusters will be different.  Simulations of these systems in three dimensions will need to be performed to gain further insight on the role of dimensionality.

It should be finally mentioned that our model of active Brownian particles ignores 
hydrodynamic effects from the solvent. Although there is experimental evidence that this does not qualitatively affect the formation of living clusters or micro-rotors, as both phases have been observed experimentally~\cite{palacci2013living, schwarz2012phase}, from a quantitative standpoint, the details of the hydrodynamic interactions between the spherical particles could affect the cluster morphology of living clusters~\cite{Navarro}. Also, long-range interactions may develop between spinning clusters of dumbbells, that could lead to cluster-cluster rotational velocity coupling and, under certain circumstances, overwhelm the long-range repulsive interactions between the particles and destabilize the crystalline mesophase of rotors. Nevertheless, we expect that judicious control of the strength of the long-range repulsive interaction between the particles against the active forces should be able to counteract such a destabilization. More work in this direction is underway.

\section*{Acknowledgments}
We thank Stewart Mallory for insightful discussions and helpful comments.  AC acknowledges financial supported from the National Science Foundation under  Grant No. DMR-1408259. CV acknowledges financial support from a Ramon y Cajal tenure track, and from the National Project FIS2013-43209-P.
\bibliographystyle{apsrev4-1}
\bibliography{mybib}

%merlin.mbs apsrev4-1.bst 2010-07-25 4.21a (PWD, AO, DPC) hacked
%Control: key (0)
%Control: author (72) initials jnrlst
%Control: editor formatted (1) identically to author
%Control: production of article title (-1) disabled
%Control: page (0) single
%Control: year (1) truncated
%Control: production of eprint (0) enabled
\begin{thebibliography}{36}%
\makeatletter
\providecommand \@ifxundefined [1]{%
 \@ifx{#1\undefined}
}%
\providecommand \@ifnum [1]{%
 \ifnum #1\expandafter \@firstoftwo
 \else \expandafter \@secondoftwo
 \fi
}%
\providecommand \@ifx [1]{%
 \ifx #1\expandafter \@firstoftwo
 \else \expandafter \@secondoftwo
 \fi
}%
\providecommand \natexlab [1]{#1}%
\providecommand \enquote  [1]{``#1''}%
\providecommand \bibnamefont  [1]{#1}%
\providecommand \bibfnamefont [1]{#1}%
\providecommand \citenamefont [1]{#1}%
\providecommand \href@noop [0]{\@secondoftwo}%
\providecommand \href [0]{\begingroup \@sanitize@url \@href}%
\providecommand \@href[1]{\@@startlink{#1}\@@href}%
\providecommand \@@href[1]{\endgroup#1\@@endlink}%
\providecommand \@sanitize@url [0]{\catcode `\\12\catcode `\$12\catcode
  `\&12\catcode `\#12\catcode `\^12\catcode `\_12\catcode `\%12\relax}%
\providecommand \@@startlink[1]{}%
\providecommand \@@endlink[0]{}%
\providecommand \url  [0]{\begingroup\@sanitize@url \@url }%
\providecommand \@url [1]{\endgroup\@href {#1}{\urlprefix }}%
\providecommand \urlprefix  [0]{URL }%
\providecommand \Eprint [0]{\href }%
\providecommand \doibase [0]{http://dx.doi.org/}%
\providecommand \selectlanguage [0]{\@gobble}%
\providecommand \bibinfo  [0]{\@secondoftwo}%
\providecommand \bibfield  [0]{\@secondoftwo}%
\providecommand \translation [1]{[#1]}%
\providecommand \BibitemOpen [0]{}%
\providecommand \bibitemStop [0]{}%
\providecommand \bibitemNoStop [0]{.\EOS\space}%
\providecommand \EOS [0]{\spacefactor3000\relax}%
\providecommand \BibitemShut  [1]{\csname bibitem#1\endcsname}%
\let\auto@bib@innerbib\@empty
%</preamble>
\bibitem [{\citenamefont {Alberts}\ \emph {et~al.}(2008)\citenamefont
  {Alberts}, \citenamefont {Bray}, \citenamefont {Lewis}, \citenamefont {Raff},
  \citenamefont {Roberts},\ and\ \citenamefont {Watson}}]{Alberts}%
  \BibitemOpen
  \bibfield  {author} {\bibinfo {author} {\bibfnamefont {B.}~\bibnamefont
  {Alberts}}, \bibinfo {author} {\bibfnamefont {D.}~\bibnamefont {Bray}},
  \bibinfo {author} {\bibfnamefont {J.}~\bibnamefont {Lewis}}, \bibinfo
  {author} {\bibfnamefont {M.}~\bibnamefont {Raff}}, \bibinfo {author}
  {\bibfnamefont {K.}~\bibnamefont {Roberts}}, \ and\ \bibinfo {author}
  {\bibfnamefont {J.~D.}\ \bibnamefont {Watson}},\ }\href@noop {} {\emph
  {\bibinfo {title} {Molecular Biology of the Cell}}}\ (\bibinfo  {publisher}
  {Garland Science},\ \bibinfo {address} {New York \& Oxford},\ \bibinfo {year}
  {2008})\BibitemShut {NoStop}%
\bibitem [{\citenamefont {Koch}\ and\ \citenamefont
  {Meinhardt}(1994)}]{Meinhardt}%
  \BibitemOpen
  \bibfield  {author} {\bibinfo {author} {\bibfnamefont {A.~J.}\ \bibnamefont
  {Koch}}\ and\ \bibinfo {author} {\bibfnamefont {H.}~\bibnamefont
  {Meinhardt}},\ }\href@noop {} {\bibfield  {journal} {\bibinfo  {journal}
  {Rev. of Modern Physics}\ }\textbf {\bibinfo {volume} {66}},\ \bibinfo
  {pages} {1481} (\bibinfo {year} {1994})}\BibitemShut {NoStop}%
\bibitem [{\citenamefont {Vedmedenko}(2007)}]{Vedmedenko}%
  \BibitemOpen
  \bibfield  {author} {\bibinfo {author} {\bibfnamefont {E.}~\bibnamefont
  {Vedmedenko}},\ }\href@noop {} {\emph {\bibinfo {title} {Competing
  Interactions and Pattern Formation in Nanoworld}}}\ (\bibinfo  {publisher}
  {WILEY-VCH},\ \bibinfo {address} {Weinheim \& Germany},\ \bibinfo {year}
  {2007})\BibitemShut {NoStop}%
\bibitem [{\citenamefont {Leibler}(1980)}]{Leibler}%
  \BibitemOpen
  \bibfield  {author} {\bibinfo {author} {\bibfnamefont {L.}~\bibnamefont
  {Leibler}},\ }\href@noop {} {\bibfield  {journal} {\bibinfo  {journal}
  {Macromolecules}\ }\textbf {\bibinfo {volume} {13}},\ \bibinfo {pages} {1602}
  (\bibinfo {year} {1980})}\BibitemShut {NoStop}%
\bibitem [{\citenamefont {Blanazs}\ \emph {et~al.}(2009)\citenamefont
  {Blanazs}, \citenamefont {Armes},\ and\ \citenamefont {Ryan}}]{Blanazs}%
  \BibitemOpen
  \bibfield  {author} {\bibinfo {author} {\bibfnamefont {A.}~\bibnamefont
  {Blanazs}}, \bibinfo {author} {\bibfnamefont {S.~P.}\ \bibnamefont {Armes}},
  \ and\ \bibinfo {author} {\bibfnamefont {A.~J.}\ \bibnamefont {Ryan}},\
  }\href@noop {} {\bibfield  {journal} {\bibinfo  {journal} {Macromol. Rapid
  Commun.}\ }\textbf {\bibinfo {volume} {30}},\ \bibinfo {pages} {267}
  (\bibinfo {year} {2009})}\BibitemShut {NoStop}%
\bibitem [{\citenamefont {Smarta}\ \emph {et~al.}(2008)\citenamefont {Smarta},
  \citenamefont {Lomasa}, \citenamefont {Massignania}, \citenamefont
  {Flores-Merinoa}, \citenamefont {Perezb},\ and\ \citenamefont
  {Battaglia}}]{battaglia}%
  \BibitemOpen
  \bibfield  {author} {\bibinfo {author} {\bibfnamefont {T.}~\bibnamefont
  {Smarta}}, \bibinfo {author} {\bibfnamefont {H.}~\bibnamefont {Lomasa}},
  \bibinfo {author} {\bibfnamefont {M.}~\bibnamefont {Massignania}}, \bibinfo
  {author} {\bibfnamefont {M.~V.}\ \bibnamefont {Flores-Merinoa}}, \bibinfo
  {author} {\bibfnamefont {L.~R.}\ \bibnamefont {Perezb}}, \ and\ \bibinfo
  {author} {\bibfnamefont {G.}~\bibnamefont {Battaglia}},\ }\href@noop {}
  {\bibfield  {journal} {\bibinfo  {journal} {Nanotoday}\ }\textbf {\bibinfo
  {volume} {3}},\ \bibinfo {pages} {38} (\bibinfo {year} {2008})}\BibitemShut
  {NoStop}%
\bibitem [{\citenamefont {Kim}\ \emph {et~al.}(2010)\citenamefont {Kim},
  \citenamefont {Park},\ and\ \citenamefont {Hinsberg}}]{Hinsberg}%
  \BibitemOpen
  \bibfield  {author} {\bibinfo {author} {\bibfnamefont {H.}~\bibnamefont
  {Kim}}, \bibinfo {author} {\bibfnamefont {S.}~\bibnamefont {Park}}, \ and\
  \bibinfo {author} {\bibfnamefont {W.~D.}\ \bibnamefont {Hinsberg}},\
  }\href@noop {} {\bibfield  {journal} {\bibinfo  {journal} {Chem. Rev.}\
  }\textbf {\bibinfo {volume} {110}},\ \bibinfo {pages} {146} (\bibinfo {year}
  {2010})}\BibitemShut {NoStop}%
\bibitem [{\citenamefont {Stradner}\ \emph {et~al.}(2004)\citenamefont
  {Stradner}, \citenamefont {Sedgwick}, \citenamefont {Cardinaux},
  \citenamefont {Poon}, \citenamefont {Egelhaaf},\ and\ \citenamefont
  {Schurtenberger}}]{poon2}%
  \BibitemOpen
  \bibfield  {author} {\bibinfo {author} {\bibfnamefont {A.}~\bibnamefont
  {Stradner}}, \bibinfo {author} {\bibfnamefont {H.}~\bibnamefont {Sedgwick}},
  \bibinfo {author} {\bibfnamefont {F.}~\bibnamefont {Cardinaux}}, \bibinfo
  {author} {\bibfnamefont {W.}~\bibnamefont {Poon}}, \bibinfo {author}
  {\bibfnamefont {S.}~\bibnamefont {Egelhaaf}}, \ and\ \bibinfo {author}
  {\bibfnamefont {P.}~\bibnamefont {Schurtenberger}},\ }\href@noop {}
  {\bibfield  {journal} {\bibinfo  {journal} {Nature}\ }\textbf {\bibinfo
  {volume} {432}},\ \bibinfo {pages} {492} (\bibinfo {year}
  {2004})}\BibitemShut {NoStop}%
\bibitem [{\citenamefont {Sciortino}\ \emph {et~al.}(2004)\citenamefont
  {Sciortino}, \citenamefont {Mossa}, \citenamefont {Zaccarelli},\ and\
  \citenamefont {Tartaglia}}]{Sciortino}%
  \BibitemOpen
  \bibfield  {author} {\bibinfo {author} {\bibfnamefont {F.}~\bibnamefont
  {Sciortino}}, \bibinfo {author} {\bibfnamefont {S.}~\bibnamefont {Mossa}},
  \bibinfo {author} {\bibfnamefont {E.}~\bibnamefont {Zaccarelli}}, \ and\
  \bibinfo {author} {\bibfnamefont {P.}~\bibnamefont {Tartaglia}},\ }\href@noop
  {} {\bibfield  {journal} {\bibinfo  {journal} {Physical Review Letters}\
  }\textbf {\bibinfo {volume} {93}},\ \bibinfo {pages} {055701} (\bibinfo
  {year} {2004})}\BibitemShut {NoStop}%
\bibitem [{\citenamefont {Imperio}\ and\ \citenamefont
  {Reatto}(2004)}]{imperio2004bidimensional}%
  \BibitemOpen
  \bibfield  {author} {\bibinfo {author} {\bibfnamefont {A.}~\bibnamefont
  {Imperio}}\ and\ \bibinfo {author} {\bibfnamefont {L.}~\bibnamefont
  {Reatto}},\ }\href@noop {} {\bibfield  {journal} {\bibinfo  {journal}
  {Journal of Physics: Condensed Matter}\ }\textbf {\bibinfo {volume} {16}},\
  \bibinfo {pages} {S3769} (\bibinfo {year} {2004})}\BibitemShut {NoStop}%
\bibitem [{\citenamefont {Imperio}\ and\ \citenamefont
  {Reatto}(2006)}]{imperio2006microphase}%
  \BibitemOpen
  \bibfield  {author} {\bibinfo {author} {\bibfnamefont {A.}~\bibnamefont
  {Imperio}}\ and\ \bibinfo {author} {\bibfnamefont {L.}~\bibnamefont
  {Reatto}},\ }\href@noop {} {\bibfield  {journal} {\bibinfo  {journal} {The
  Journal of Chemical Physics}\ }\textbf {\bibinfo {volume} {124}},\ \bibinfo
  {pages} {164712} (\bibinfo {year} {2006})}\BibitemShut {NoStop}%
\bibitem [{\citenamefont {Imperio}\ and\ \citenamefont
  {Reatto}(2007)}]{imperio2007microphase}%
  \BibitemOpen
  \bibfield  {author} {\bibinfo {author} {\bibfnamefont {A.}~\bibnamefont
  {Imperio}}\ and\ \bibinfo {author} {\bibfnamefont {L.}~\bibnamefont
  {Reatto}},\ }\href@noop {} {\bibfield  {journal} {\bibinfo  {journal}
  {Physical Review E}\ }\textbf {\bibinfo {volume} {76}},\ \bibinfo {pages}
  {040402} (\bibinfo {year} {2007})}\BibitemShut {NoStop}%
\bibitem [{\citenamefont {Imperio}\ \emph {et~al.}(2008)\citenamefont
  {Imperio}, \citenamefont {Reatto},\ and\ \citenamefont
  {Zapperi}}]{imperio2008rheology}%
  \BibitemOpen
  \bibfield  {author} {\bibinfo {author} {\bibfnamefont {A.}~\bibnamefont
  {Imperio}}, \bibinfo {author} {\bibfnamefont {L.}~\bibnamefont {Reatto}}, \
  and\ \bibinfo {author} {\bibfnamefont {S.}~\bibnamefont {Zapperi}},\
  }\href@noop {} {\bibfield  {journal} {\bibinfo  {journal} {Physical Review
  E}\ }\textbf {\bibinfo {volume} {78}},\ \bibinfo {pages} {021402} (\bibinfo
  {year} {2008})}\BibitemShut {NoStop}%
\bibitem [{\citenamefont {De’Bell}\ \emph {et~al.}(2000)\citenamefont
  {De’Bell}, \citenamefont {MacIsaac},\ and\ \citenamefont
  {Whitehead}}]{de2000dipolar}%
  \BibitemOpen
  \bibfield  {author} {\bibinfo {author} {\bibfnamefont {K.}~\bibnamefont
  {De’Bell}}, \bibinfo {author} {\bibfnamefont {A.}~\bibnamefont {MacIsaac}},
  \ and\ \bibinfo {author} {\bibfnamefont {J.}~\bibnamefont {Whitehead}},\
  }\href@noop {} {\bibfield  {journal} {\bibinfo  {journal} {Reviews of Modern
  Physics}\ }\textbf {\bibinfo {volume} {72}},\ \bibinfo {pages} {225}
  (\bibinfo {year} {2000})}\BibitemShut {NoStop}%
\bibitem [{\citenamefont {Muratov}(2002)}]{muratov2002theory}%
  \BibitemOpen
  \bibfield  {author} {\bibinfo {author} {\bibfnamefont {C.}~\bibnamefont
  {Muratov}},\ }\href@noop {} {\bibfield  {journal} {\bibinfo  {journal}
  {Physical Review E}\ }\textbf {\bibinfo {volume} {66}},\ \bibinfo {pages}
  {066108} (\bibinfo {year} {2002})}\BibitemShut {NoStop}%
\bibitem [{\citenamefont {Ghezzi}\ and\ \citenamefont
  {Earnshaw}(1997)}]{ghezzi1997formation}%
  \BibitemOpen
  \bibfield  {author} {\bibinfo {author} {\bibfnamefont {F.}~\bibnamefont
  {Ghezzi}}\ and\ \bibinfo {author} {\bibfnamefont {J.}~\bibnamefont
  {Earnshaw}},\ }\href@noop {} {\bibfield  {journal} {\bibinfo  {journal}
  {Journal of Physics: Condensed Matter}\ }\textbf {\bibinfo {volume} {9}},\
  \bibinfo {pages} {L517} (\bibinfo {year} {1997})}\BibitemShut {NoStop}%
\bibitem [{\citenamefont {Ghezzi}\ \emph {et~al.}(2001)\citenamefont {Ghezzi},
  \citenamefont {Earnshaw}, \citenamefont {Finnis},\ and\ \citenamefont
  {McCluney}}]{ghezzi2001pattern}%
  \BibitemOpen
  \bibfield  {author} {\bibinfo {author} {\bibfnamefont {F.}~\bibnamefont
  {Ghezzi}}, \bibinfo {author} {\bibfnamefont {J.}~\bibnamefont {Earnshaw}},
  \bibinfo {author} {\bibfnamefont {M.}~\bibnamefont {Finnis}}, \ and\ \bibinfo
  {author} {\bibfnamefont {M.}~\bibnamefont {McCluney}},\ }\href@noop {}
  {\bibfield  {journal} {\bibinfo  {journal} {Journal of Colloid and Interface
  Science}\ }\textbf {\bibinfo {volume} {238}},\ \bibinfo {pages} {433}
  (\bibinfo {year} {2001})}\BibitemShut {NoStop}%
\bibitem [{\citenamefont {Palacci}\ \emph {et~al.}(2013)\citenamefont
  {Palacci}, \citenamefont {Sacanna}, \citenamefont {Steinberg}, \citenamefont
  {Pine},\ and\ \citenamefont {Chaikin}}]{palacci2013living}%
  \BibitemOpen
  \bibfield  {author} {\bibinfo {author} {\bibfnamefont {J.}~\bibnamefont
  {Palacci}}, \bibinfo {author} {\bibfnamefont {S.}~\bibnamefont {Sacanna}},
  \bibinfo {author} {\bibfnamefont {A.~P.}\ \bibnamefont {Steinberg}}, \bibinfo
  {author} {\bibfnamefont {D.~J.}\ \bibnamefont {Pine}}, \ and\ \bibinfo
  {author} {\bibfnamefont {P.~M.}\ \bibnamefont {Chaikin}},\ }\href@noop {}
  {\bibfield  {journal} {\bibinfo  {journal} {Science}\ }\textbf {\bibinfo
  {volume} {339}},\ \bibinfo {pages} {936} (\bibinfo {year}
  {2013})}\BibitemShut {NoStop}%
\bibitem [{\citenamefont {Theurkoff}\ \emph {et~al.}(2012)\citenamefont
  {Theurkoff}, \citenamefont {Cottin-Bizonne}, \citenamefont {Palacci},
  \citenamefont {Ybert},\ and\ \citenamefont {Bocquet}}]{theurkauff}%
  \BibitemOpen
  \bibfield  {author} {\bibinfo {author} {\bibfnamefont {I.}~\bibnamefont
  {Theurkoff}}, \bibinfo {author} {\bibfnamefont {C.}~\bibnamefont
  {Cottin-Bizonne}}, \bibinfo {author} {\bibfnamefont {J.}~\bibnamefont
  {Palacci}}, \bibinfo {author} {\bibfnamefont {C.}~\bibnamefont {Ybert}}, \
  and\ \bibinfo {author} {\bibfnamefont {L.}~\bibnamefont {Bocquet}},\
  }\href@noop {} {\bibfield  {journal} {\bibinfo  {journal} {Physical Review
  Letters}\ }\textbf {\bibinfo {volume} {108}},\ \bibinfo {pages} {268303}
  (\bibinfo {year} {2012})}\BibitemShut {NoStop}%
\bibitem [{\citenamefont {Buttinoni}\ \emph {et~al.}(2013)\citenamefont
  {Buttinoni}, \citenamefont {Bialk{\'e}}, \citenamefont {K{\"u}mmel},
  \citenamefont {L{\"o}wen}, \citenamefont {Bechinger},\ and\ \citenamefont
  {Speck}}]{buttinoni2013dynamical}%
  \BibitemOpen
  \bibfield  {author} {\bibinfo {author} {\bibfnamefont {I.}~\bibnamefont
  {Buttinoni}}, \bibinfo {author} {\bibfnamefont {J.}~\bibnamefont
  {Bialk{\'e}}}, \bibinfo {author} {\bibfnamefont {F.}~\bibnamefont
  {K{\"u}mmel}}, \bibinfo {author} {\bibfnamefont {H.}~\bibnamefont
  {L{\"o}wen}}, \bibinfo {author} {\bibfnamefont {C.}~\bibnamefont
  {Bechinger}}, \ and\ \bibinfo {author} {\bibfnamefont {T.}~\bibnamefont
  {Speck}},\ }\href@noop {} {\bibfield  {journal} {\bibinfo  {journal}
  {Physical Review letters}\ }\textbf {\bibinfo {volume} {110}},\ \bibinfo
  {pages} {238301} (\bibinfo {year} {2013})}\BibitemShut {NoStop}%
\bibitem [{\citenamefont {Mognetti}\ \emph {et~al.}(2013)\citenamefont
  {Mognetti}, \citenamefont {{\v{S}}ari{\'c}}, \citenamefont
  {Angioletti-Uberti}, \citenamefont {Cacciuto}, \citenamefont {Valeriani},\
  and\ \citenamefont {Frenkel}}]{mognetti2013living}%
  \BibitemOpen
  \bibfield  {author} {\bibinfo {author} {\bibfnamefont {B.~M.}\ \bibnamefont
  {Mognetti}}, \bibinfo {author} {\bibfnamefont {A.}~\bibnamefont
  {{\v{S}}ari{\'c}}}, \bibinfo {author} {\bibfnamefont {S.}~\bibnamefont
  {Angioletti-Uberti}}, \bibinfo {author} {\bibfnamefont {A.}~\bibnamefont
  {Cacciuto}}, \bibinfo {author} {\bibfnamefont {C.}~\bibnamefont {Valeriani}},
  \ and\ \bibinfo {author} {\bibfnamefont {D.}~\bibnamefont {Frenkel}},\
  }\href@noop {} {\bibfield  {journal} {\bibinfo  {journal} {Physical Review
  Letters}\ }\textbf {\bibinfo {volume} {111}},\ \bibinfo {pages} {245702}
  (\bibinfo {year} {2013})}\BibitemShut {NoStop}%
\bibitem [{\citenamefont {Prymidis}\ \emph {et~al.}(2015)\citenamefont
  {Prymidis}, \citenamefont {Sielcken},\ and\ \citenamefont {Filion}}]{filion}%
  \BibitemOpen
  \bibfield  {author} {\bibinfo {author} {\bibfnamefont {V.}~\bibnamefont
  {Prymidis}}, \bibinfo {author} {\bibfnamefont {H.}~\bibnamefont {Sielcken}},
  \ and\ \bibinfo {author} {\bibfnamefont {L.}~\bibnamefont {Filion}},\
  }\href@noop {} {\bibfield  {journal} {\bibinfo  {journal} {Soft Matter}\
  }\textbf {\bibinfo {volume} {11}},\ \bibinfo {pages} {4158} (\bibinfo {year}
  {2015})}\BibitemShut {NoStop}%
\bibitem [{\citenamefont {Schwarz-Linek}\ \emph {et~al.}(2012)\citenamefont
  {Schwarz-Linek}, \citenamefont {Valeriani}, \citenamefont {Cacciuto},
  \citenamefont {Cates}, \citenamefont {Marenduzzo}, \citenamefont {Morozov},\
  and\ \citenamefont {Poon}}]{schwarz2012phase}%
  \BibitemOpen
  \bibfield  {author} {\bibinfo {author} {\bibfnamefont {J.}~\bibnamefont
  {Schwarz-Linek}}, \bibinfo {author} {\bibfnamefont {C.}~\bibnamefont
  {Valeriani}}, \bibinfo {author} {\bibfnamefont {A.}~\bibnamefont {Cacciuto}},
  \bibinfo {author} {\bibfnamefont {M.}~\bibnamefont {Cates}}, \bibinfo
  {author} {\bibfnamefont {D.}~\bibnamefont {Marenduzzo}}, \bibinfo {author}
  {\bibfnamefont {A.}~\bibnamefont {Morozov}}, \ and\ \bibinfo {author}
  {\bibfnamefont {W.}~\bibnamefont {Poon}},\ }\href@noop {} {\bibfield
  {journal} {\bibinfo  {journal} {PNAS}\ }\textbf {\bibinfo {volume} {109}},\
  \bibinfo {pages} {4052} (\bibinfo {year} {2012})}\BibitemShut {NoStop}%
\bibitem [{\citenamefont {Suma}\ \emph {et~al.}(2014)\citenamefont {Suma},
  \citenamefont {Gonnella}, \citenamefont {Marenduzzo},\ and\ \citenamefont
  {Orlandini}}]{maren}%
  \BibitemOpen
  \bibfield  {author} {\bibinfo {author} {\bibfnamefont {A.}~\bibnamefont
  {Suma}}, \bibinfo {author} {\bibfnamefont {G.}~\bibnamefont {Gonnella}},
  \bibinfo {author} {\bibfnamefont {D.}~\bibnamefont {Marenduzzo}}, \ and\
  \bibinfo {author} {\bibfnamefont {E.}~\bibnamefont {Orlandini}},\ }\href@noop
  {} {\bibfield  {journal} {\bibinfo  {journal} {E.P.L.}\ }\textbf {\bibinfo
  {volume} {108}},\ \bibinfo {pages} {56004} (\bibinfo {year}
  {2014})}\BibitemShut {NoStop}%
\bibitem [{\citenamefont {Bal}(2013)}]{PRXX}%
  \BibitemOpen
  \bibfield  {author} {\bibinfo {author} {\bibfnamefont {P.}~\bibnamefont
  {Bal}},\ }\href@noop {} {\bibfield  {journal} {\bibinfo  {journal} {Physics}\
  }\textbf {\bibinfo {volume} {6}},\ \bibinfo {pages} {134} (\bibinfo {year}
  {2013})}\BibitemShut {NoStop}%
\bibitem [{\citenamefont {Heptner}\ and\ \citenamefont
  {Dzubiella}(2015)}]{heptner2015equilibrium}%
  \BibitemOpen
  \bibfield  {author} {\bibinfo {author} {\bibfnamefont {N.}~\bibnamefont
  {Heptner}}\ and\ \bibinfo {author} {\bibfnamefont {J.}~\bibnamefont
  {Dzubiella}},\ }\href@noop {} {\bibfield  {journal} {\bibinfo  {journal}
  {Molecular Physics}\ }\textbf {\bibinfo {volume} {113}},\ \bibinfo {pages}
  {1} (\bibinfo {year} {2015})}\BibitemShut {NoStop}%
\bibitem [{\citenamefont {Kirkwood}\ and\ \citenamefont
  {Riseman}(1948)}]{kirkwood}%
  \BibitemOpen
  \bibfield  {author} {\bibinfo {author} {\bibfnamefont {J.}~\bibnamefont
  {Kirkwood}}\ and\ \bibinfo {author} {\bibfnamefont {J.}~\bibnamefont
  {Riseman}},\ }\href@noop {} {\bibfield  {journal} {\bibinfo  {journal} {J.
  Chem. Phys.}\ }\textbf {\bibinfo {volume} {16}},\ \bibinfo {pages} {565}
  (\bibinfo {year} {1948})}\BibitemShut {NoStop}%
\bibitem [{\citenamefont {Moreno}\ and\ \citenamefont
  {Likos}(2007)}]{moreno2007diffusion}%
  \BibitemOpen
  \bibfield  {author} {\bibinfo {author} {\bibfnamefont {A.~J.}\ \bibnamefont
  {Moreno}}\ and\ \bibinfo {author} {\bibfnamefont {C.~N.}\ \bibnamefont
  {Likos}},\ }\href@noop {} {\bibfield  {journal} {\bibinfo  {journal}
  {Physical Review Letters}\ }\textbf {\bibinfo {volume} {99}},\ \bibinfo
  {pages} {107801} (\bibinfo {year} {2007})}\BibitemShut {NoStop}%
\bibitem [{\citenamefont {Mladek}\ \emph {et~al.}(2007)\citenamefont {Mladek},
  \citenamefont {Charbonneau},\ and\ \citenamefont {Frenkel}}]{mladekk}%
  \BibitemOpen
  \bibfield  {author} {\bibinfo {author} {\bibfnamefont {B.~M.}\ \bibnamefont
  {Mladek}}, \bibinfo {author} {\bibfnamefont {P.}~\bibnamefont {Charbonneau}},
  \ and\ \bibinfo {author} {\bibfnamefont {D.}~\bibnamefont {Frenkel}},\
  }\href@noop {} {\bibfield  {journal} {\bibinfo  {journal} {Physical Revew
  Letters}\ }\textbf {\bibinfo {volume} {99}},\ \bibinfo {pages} {235702}
  (\bibinfo {year} {2007})}\BibitemShut {NoStop}%
\bibitem [{\citenamefont {Nelson}\ and\ \citenamefont
  {Halperin}(1979)}]{nelson1979dislocation}%
  \BibitemOpen
  \bibfield  {author} {\bibinfo {author} {\bibfnamefont {D.~R.}\ \bibnamefont
  {Nelson}}\ and\ \bibinfo {author} {\bibfnamefont {B.}~\bibnamefont
  {Halperin}},\ }\href@noop {} {\bibfield  {journal} {\bibinfo  {journal}
  {Physical Review B}\ }\textbf {\bibinfo {volume} {19}},\ \bibinfo {pages}
  {2457} (\bibinfo {year} {1979})}\BibitemShut {NoStop}%
\bibitem [{\citenamefont {Cates}\ and\ \citenamefont {Tailleur}(2013)}]{sp1}%
  \BibitemOpen
  \bibfield  {author} {\bibinfo {author} {\bibfnamefont {M.~E.}\ \bibnamefont
  {Cates}}\ and\ \bibinfo {author} {\bibfnamefont {J.}~\bibnamefont
  {Tailleur}},\ }\href@noop {} {\bibfield  {journal} {\bibinfo  {journal} {E.
  P. L.}\ }\textbf {\bibinfo {volume} {101}},\ \bibinfo {pages} {20010}
  (\bibinfo {year} {2013})}\BibitemShut {NoStop}%
\bibitem [{\citenamefont {Fily}\ and\ \citenamefont {Marchetti}(2012)}]{sp2}%
  \BibitemOpen
  \bibfield  {author} {\bibinfo {author} {\bibfnamefont {Y.}~\bibnamefont
  {Fily}}\ and\ \bibinfo {author} {\bibfnamefont {M.~C.}\ \bibnamefont
  {Marchetti}},\ }\href@noop {} {\bibfield  {journal} {\bibinfo  {journal}
  {Physical Review Letters}\ }\textbf {\bibinfo {volume} {108}},\ \bibinfo
  {pages} {235702} (\bibinfo {year} {2012})}\BibitemShut {NoStop}%
\bibitem [{\citenamefont {Redner}\ \emph
  {et~al.}(2013{\natexlab{a}})\citenamefont {Redner}, \citenamefont {Hagan},\
  and\ \citenamefont {Baskaran}}]{sp3}%
  \BibitemOpen
  \bibfield  {author} {\bibinfo {author} {\bibfnamefont {G.~S.}\ \bibnamefont
  {Redner}}, \bibinfo {author} {\bibfnamefont {M.~F.}\ \bibnamefont {Hagan}}, \
  and\ \bibinfo {author} {\bibfnamefont {A.}~\bibnamefont {Baskaran}},\
  }\href@noop {} {\bibfield  {journal} {\bibinfo  {journal} {Physical Review
  Letters}\ }\textbf {\bibinfo {volume} {110}},\ \bibinfo {pages} {055701}
  (\bibinfo {year} {2013}{\natexlab{a}})}\BibitemShut {NoStop}%
\bibitem [{\citenamefont {Redner}\ \emph
  {et~al.}(2013{\natexlab{b}})\citenamefont {Redner}, \citenamefont
  {Baskaran},\ and\ \citenamefont {Hagan}}]{redner}%
  \BibitemOpen
  \bibfield  {author} {\bibinfo {author} {\bibfnamefont {G.}~\bibnamefont
  {Redner}}, \bibinfo {author} {\bibfnamefont {A.}~\bibnamefont {Baskaran}}, \
  and\ \bibinfo {author} {\bibfnamefont {M.}~\bibnamefont {Hagan}},\
  }\href@noop {} {\bibfield  {journal} {\bibinfo  {journal} {Physical Review
  E}\ }\textbf {\bibinfo {volume} {88}},\ \bibinfo {pages} {012305} (\bibinfo
  {year} {2013}{\natexlab{b}})}\BibitemShut {NoStop}%
\bibitem [{\citenamefont {Mani}\ and\ \citenamefont
  {L{\"o}wen}(2015)}]{mani2015effect}%
  \BibitemOpen
  \bibfield  {author} {\bibinfo {author} {\bibfnamefont {E.}~\bibnamefont
  {Mani}}\ and\ \bibinfo {author} {\bibfnamefont {H.}~\bibnamefont
  {L{\"o}wen}},\ }\href@noop {} {\bibfield  {journal} {\bibinfo  {journal}
  {Physical Review E}\ }\textbf {\bibinfo {volume} {92}},\ \bibinfo {pages}
  {032301} (\bibinfo {year} {2015})}\BibitemShut {NoStop}%
\bibitem [{\citenamefont {Navarro}\ and\ \citenamefont
  {Fielding}(2015)}]{Navarro}%
  \BibitemOpen
  \bibfield  {author} {\bibinfo {author} {\bibfnamefont {R.~M.}\ \bibnamefont
  {Navarro}}\ and\ \bibinfo {author} {\bibfnamefont {S.~M.}\ \bibnamefont
  {Fielding}},\ }\href@noop {} {\bibfield  {journal} {\bibinfo  {journal} {Soft
  Matter}\ }\textbf {\bibinfo {volume} {11}},\ \bibinfo {pages} {7525}
  (\bibinfo {year} {2015})}\BibitemShut {NoStop}%
\end{thebibliography}%

\newpage

% NOTE: use package adjustbox to align the minipages, for later...
%  

\end{document}